\newcommand{\gev}{\, {\rm GeV}}
\newcommand{\tev}{\, {\rm TeV}}

\newcommand{\pev}{\, {\rm PeV}}
\newcommand{\beq}{\begin{equation}}
\newcommand{\eeq}{\end{equation}}
\newcommand{\bea}{\begin{eqnarray}}
\newcommand{\eea}{\end{eqnarray}}

\newcommand{\gsim}{\lower.7ex\hbox{$\;\stackrel{\textstyle>}{\sim}\;$}}
\newcommand{\lsim}{\lower.7ex\hbox{$\;\stackrel{\textstyle<}{\sim}\;$}}

\documentclass[aps,prl,twocolumn,preprintnumbers,%
               floatfix,nofootinbib]{revtex4}
\usepackage{graphicx}

\def\stacksymbols #1#2#3#4{\def\theguybelow{#2}
    \def\vp{\lower#3pt}
    \def\sp{\baselineskip0pt\lineskip#4pt}
    \mathrel{\mathpalette\intermediary#1}}

\def\intermediary#1#2{\vp\vbox{\sp
     \everycr={}\tabskip0pt
     \halign{$\mathsurround0pt#1\hfil##\hfil$\crcr#2\crcr
              \theguybelow\crcr}}}

\def\to{\rightarrow}

\begin{document}

%
%

\preprint{MCTP-04-61, hep-ph/0411041}

\title{PeV-Scale Supersymmetry}

\author{James D. Wells} 
\vspace{0.2cm}
\affiliation{
Michigan Center for Theoretical Physics (MCTP) \\
Department of Physics, University of Michigan, Ann Arbor, MI 48109}

\begin{abstract}

Although supersymmetry has not been seen directly by experiment,
there are
powerful physics reasons to suspect that it should be an ingredient of nature
and that superpartner masses should be somewhat near the weak scale.  
I present an argument that if we dismiss our ordinary intuition of finetuning,
and focus entirely on more concrete physics issues, the PeV scale 
might be the best place for supersymmetry.
PeV-scale supersymmetry admits gauge coupling unification, predicts a Higgs mass between
125 GeV and 155 GeV, and generally
disallows flavor changing neutral currents and CP violating effects in conflict with current experiment.   The PeV scale is motivated independently by dark matter and neutrino mass considerations.

\end{abstract}

\maketitle


\maketitle


\setcounter{equation}{0}
{\bf Introduction:}
Finetuning arguments are not vacuous.  The explanations for a finetuned
scenario, such as the weak-to-Planck scale hierarchy, may either be
based on a forcing or stabilization principle (Principled Finetuning) or 
based on accrued probability from a large ensemble of possibilities with small random chance
(Chance Finetuning).

The electroweak symmetry breaking potential in supersymmetry involves superpartner
mass terms that must conspire to ultimately give the $Z$ scale through the 
electroweak symmetry breaking condition
\beq
\frac{1}{2}m_Z^2 +\mu^2= \frac{m^2_{H_d} -m^2_{H_u}\tan^2\beta}{\tan^2\beta-1}.
\eeq
We are normally committed to explaining the
electroweak scale through Principled Finetuning arguments.
The masses $m^2_{H_u}$, $m^2_{H_d}$ and $\mu^2$ can be obtained by a natural
dimensional transmutation associated with supersymmetry breaking.
These masses should be near the $Z$ scale for there to be no finetuning in
the electroweak potential.  This viewpoint is what has dominated physics
discussions for many years, and in particular the discussion surrounding supersymmetry.

Another viewpoint is that Chance Finetuning explains the 
weak scale.  Analysis of the string/M-theory landscape indicates that there
is a {\it plausible} argument that we may need to dismiss Principled Finetuning
as a criterion and apply Chance Finetuning to some key features of 
nature~\cite{Susskind:2004uv}.
This idea gains most traction when trying to explain the smallness of the 
cosmological constant, which might have a more attractive
Chance Finetuning explanation than Principled Finetuning
explanation~\cite{Bousso:2000xa}. 
The counting of vacua on the string/M-theory landscape can
be equally applied to the related issue of the hierarchy.

The point of the landscape is that overall numbers of vacua for one configuration
(high-scale susy, small cosmological constant) might so overwhelm the number
of vacua of another configuration (low-scale susy, small cosmological constant),
that it is plausible that Principled Finetuning arguments applied to the hierarchy
problem are not wholly relevant.  Given the interesting results coming from string/M-theory
vacua counting, and the long-standing challenges of quantifying what
finetuning really means, it is defensible to make the following claim:

{\it In our present state of understanding, we cannot determine if the finetuning associated with
the weak-to-Planck scale hierarchy is solved by a Principled Finetuning
argument or by a Chance Finetuning argument.}

Therefore, if the above statement is agreed to, one is free to
dismiss weak-scale Principled Finetuning arguments as a guide
to model building, and investigate the consequences.  
This idea reaches its most intense expression in split 
supersymmetry~\cite{Arkani-Hamed:2004fb}-\cite{Arkani-Hamed:2004yi},
where a dramatic separation between gauginos and scalar superpartners
are possible. I will follow a smaller subset of this general view, 
building from a conference discussion~\cite{Wells:2003tf}, 
which one could call loop-split supersymmetry,
as the gauginos and scalar masses will be split by only a loop factor associated
with anomaly mediation.  Once we truly dismiss Principled Finetuning
considerations, keeping an eye toward generic supersymmetry breaking
mechanisms that are compatible with the data, I suggest that we are
drawn to a scenario where the supersymmetry breaking mass (i.e., gravitino mass)
is at the PeV scale (${\rm PeV}=10^{15}\, {\rm eV}$).

{\bf Data Pressures on Supersymmetry:}
First, we briefly review the negative pressures data has placed
on supersymmetry. Results from flavor
changing neutral current experiments ($K-\bar K$ mixing, $\mu\to e\gamma$,
etc.), CP violation experiments (e.g., electric dipole moments of the neutron
and electron), and Higgs mass searches ($m_h>114\gev$ at 95\% C.L.)
all struggle to be consistent with weak-scale 
supersymmetry~\cite{Chung:2003fi}.  One must make
additional assumptions about the superpartner spectrum, such
as the squarks must be degenerate and CP phases of superpartner
parameters ($\mu$, gaugino masses, $A$-terms) must be nearly zero.
Proton decay is another potential problem. Proposed solutions based on natural
R-parity arguments mollify dimension-four concerns, but dimension-five
operators still frighten the grand unified theory enthusiasts.

If scalar superpartner masses are all above a few hundred TeV the problems
discussed above are solved.
Predictions of flavor changing neutral currents in 
Kaon physics and B physics are identical to the predictions of 
the Standard Model if the scalar superpartners are in the PeV range.
The experiments show no compelling deviations from the
Standard Model predictions, and so PeV scale superpartners work.
Unwanted CP violating effects are also no longer present.  
(There is a two-loop
contribution to dipole moments that could be accessible at
experiments in the near future if gauginos and Higgsinos are
light~\cite{Arkani-Hamed:2004fb}, but we will not be pursuing
that direction below.)
Furthermore,
as the squark masses increase, the troubling dimension-five proton
decay operators are suppressed and proton decay is much less of
a concern~\cite{Arnowitt:1998uz}. 

There is a dark matter concern which needs to be addressed
when the scalar superpartner masses get too heavy.
In ordinary supergravity models with
a Bino lightest supersymmetric particle (LSP), the relic abundance increases
to unacceptable levels in much of parameter space when the scalar
masses are increased.  This is because the Bino annihilates most 
efficiently through $t$-channel sleptons, but when those masses are
too high the annihilation efficiency drops and the relic abundance
climbs very high such that the universe is matter dominated
too early.  We will see shortly that the supersymmetry spectrum
we will be led to in this letter gives the Wino  
the honor of being the LSP.  The Wino LSP annihilates
very efficiently through ordinary gauge bosons and so the masses of
the scalars are mostly irrelevant to dark matter issues.  
More will be said about dark matter below.

Gauge coupling unification is a tantalizing indication that a higher
unification of the Standard Model forces can be accomplished
within supersymmetry.  As has been 
emphasized in~\cite{Arkani-Hamed:2004fb}, low-scale
fermionic superpartners, not the scalar superpartners, 
are what are responsible for this amazing coming together
of the gauge couplings to within a fraction of a percent, well within
the tolerances that a high-scale theory with its own threshold
corrections would need for unification.  The pressures above,
which ask for TeV scale gauginos and much heavier scalar
superpartners, are fully consistent with gauge coupling unification.

{\bf Theory Following from Data:}
Gauge coupling unification likes gauginos well
below the GUT scale, dark matter likes gauginos (or higgsinos) below several TeV,
flavor changing neutral current constraints and CP violation constraints like
scalar superpartner masses well above the tens of TeV scale, and the lightest
Higgs boson mass constraint likes scalar superpartners well above the TeV scale.

The major tension in the data is the desire for TeV gauginos and substantially
heavier scalars. Fortunately,
charged supersymmetry breaking naturally accommodates such
a tension.  By ``charged" I mean that 
there is no singlet to feel and transmit
supersymmetry breaking. Supersymmetry breaking can be parametrized
by a chiral supermultiplet $S=S+\sqrt{2}\psi\theta +F_S\theta^2$
whose non-zero $F_S$ component 
is the source of supersymmetry breaking.
Gaugino masses are generated via
\beq
\label{gaugino mass}
\int d^2\theta \frac{S}{M_{\rm Pl}}{\cal W}{\cal W}\to
 \frac{F_S}{M_{\rm Pl}}\lambda\lambda .
\eeq
The scalar masses are generated by
\beq
\label{scalar mass}
\int d^2\theta d^2\bar\theta \frac{S^\dagger S}{M_{\rm Pl}^2}\Phi^\dagger_i
\Phi_i \to \frac{F_S^\dagger F_S}{M^2_{\rm Pl}}\phi^*_i\phi_i .
\eeq
Thus, the gauginos and scalars are often of similar mass 
when considering usual supersymmetry breaking scenarios.

If $S$ is charged (i.e., not a singlet), eq.~\ref{scalar mass} is unaffected,
whereas eq.~\ref{gaugino mass} is no longer gauge invariant.
(I am neglecting the grand unified theory possibility that a representation
of $S$ charged under
the unified group paired with that of the ${\rm Adj}^2$ contains a 
singlet~\cite{Anderson:1996bg}.) This is the generic expectation in
dynamical supersymmetry breaking where supersymmetry breaking order
parameters are charged and singlets are hard to come 
by~\cite{DSB singlets}.

In this case the leading-order contribution to the
gaugino mass is the anomaly-mediated 
value~\cite{Randall:1998uk,Giudice:1998xp},
\beq
M_\lambda=\frac{\beta(g_\lambda)}{g_\lambda}m_{3/2}
\eeq
where $\lambda$ labels the three SM gauge groups, and
where $m^2_{3/2}=\langle F^\dagger_SF_S\rangle/M^2_{\rm Pl}$.  The
gaugino masses are therefore one-loop 
suppressed compared to the persisting scalar mass
result of eq.~\ref{scalar mass}.  Some phenomenological implications
of this anomaly-mediated scenario with heavier squark masses were
presented in~\cite{Giudice:1998xp}.

Charged supersymmetry breaking therefore generically creates a one-loop hierarchy
between the gaugino masses (and $A$ terms)
and the scalar superpartners. 
To lowest order, the numerical values
of the light gaugino spectrum are
\bea
M_1 & \simeq & m_{3/2}/120 \\
M_2 & \simeq & m_{3/2}/360 \\
M_3 & \simeq & m_{3/2}/40.
\eea
As discussed above, 
the heavy superpartner spectrum of squark, slepton and sneutrino
masses $\tilde m_i$ generically should 
have masses within factors of ${\cal O}(1)$ near the gravitino
mass $m_{3/2}$,
\beq
\tilde m_i \sim m_{3/2}~~({\rm scalar~masses}).  
\eeq
Thus the scalar masses are several hundred times more massive
than the lightest Wino mass.

{\bf PeV-Scale Supersymmetry from Dark Matter:}
In ordinary minimal supergravity the lightest superpartner is the 
bino, superpartner of the hypercharge gauge boson. The 
thermal relic abundance of this sparticle
can be made compatible with the universe's cold dark matter needs
in sizeable regions of the parameter space.
Weak-scale supersymmetry generally has
a gravitino and moduli problem though.  The gravitino, which is
roughly the same mass as the LSP in the usual case, decays during big bang nucleosynthesis
if its mass is less than a few TeV.
The gravitino and moduli must
be inflated away and not regenerated too copiously during the reheat
phase in this scenario. 

However, in our situation with anomaly-mediated gaugino masses, the Wino
is the LSP\footnote{For definiteness, I will assume only that $|\mu|$ is heavier than $M_2$
such that the Higgsino mixing has little effect on the thermal relic abundance of the LSP.
If $|\mu|<M_2$, which is probably not generically expected in this framework, thermal
relic abundance would still put the LSP in the TeV range.}.
The Winos annihilate and co-annihilate
very efficiently through SM gauge bosons.  Furthermore,
there is no gravitino/moduli problem as their masses should be well above
the problematic range. The thermal
relic abundance of the Winos is~\cite{Giudice:2004tc} 
\beq
\Omega^{\rm th}_{\tilde W}h^2\simeq 0.02 \left( \frac{M_2}{1\tev}\right)^2 
\label{thermal abundance}
\eeq
and is cosmologically insignificant for weak-scale gauginos, but
cosmologically interesting for Winos with mass above the TeV scale.

The WMAP experiment has analyzed the cosmic microwave 
background data~\cite{Spergel:2003cb}
at high precision. One infers from these results that the cold dark matter of the universe
should have a relic abundance
\beq
\Omega_{CDM}h^2 = 0.11\pm 0.01~~~{\rm (WMAP}~68\%~{\rm C.L.).}
\eeq
Using eq.~\ref{thermal abundance} we find that the Wino can explain the
cold dark matter of the universe if its mass is
\beq
M_{\tilde W}\simeq 2.3 \pm 0.2\tev~~{\rm (Wino~dark~matter).}
\eeq
Of course, there could be other sources of cold dark matter beyond the LSP
of supersymmetry. In that case, the above equation would set the upper
limit on the Wino mass in an R-parity conserving theory.

Detecting TeV scale Wino dark matter is a severe challenge.  When squarks and the $\mu$-term are 
in the hundreds of TeV range, detection is not possible with table-top detectors of LSP-nucleon
scattering. The coherent scattering cross-section 
falls like $1/\mu^2$. In other words,
a Higgsino component of the LSP 
is necessary to be sensitive to LSP-Nucleon
interactions, and if the LSP is nearly pure Wino the Higgsino
component is not available for service.  
The spin-dependent contribution also goes to zero, and
the sfermion contributions to the scattering go to zero as well.
Therefore, the dark matter may be invisible to 
table-top experiments.

However, Winos annihilate very efficiently and so one expects that all
experiments looking for  LSP annihilations in the galactic halo
would have an enhanced sensitivity.  For example, annihilations that 
produce $\bar p$'s and $e^+$'s 
are enhanced.  The annihilation channel that perhaps gains the most
if nature has Wino dark matter is the monochromatic 
two-photon final state~\cite{Wells:Pheno00,Ullio:2001qk}.  
The wino annihilation rate is even higher than the higgsino rate, which
is known to be large.  The 
cross-section for Winos annihilating into two photons~\cite{Bergstrom:1997fh} 
is a fairly constant value
\beq
2\sigma v(\gamma\gamma)
= (3-5)\times 10^{-27}\, {\rm cm}^{3}\, {\rm s}^{-1}
\eeq
for $m_{\tilde W}=0.1\tev - 1\pev$.

The virialized dark matter is moving at non-relativistic speeds of
only a few hundred kilometers per second, and so the photons that
result from this annihilation are monochromatic with energy
$E_\gamma=m_{\tilde W}$.  Under some astrophysical models developed
independently of dark-matter detection prospecting, 
next generation Cerenkov detectors
may be able to see a signal for $\tilde W\tilde W\to \gamma\gamma$ in
the galactic halo if the dark matter density profile is favorably
clumped near the galactic center~\cite{Ullio:2001qk}.  Another photon line from
annihilations into $Z\gamma$ might also be detectable at the energy
\beq
E_\gamma =m_{\tilde W}\left( 1-\frac{m_Z^2}{4m_{\tilde W}^2}\right)
~~({\rm from}~\tilde W\tilde W\to Z\gamma).
\eeq
which for sufficiently massive $m_{\tilde W}$ is not experimentally
resolvable in energy from the photons that come from the $\gamma\gamma$
final state.  The separation in energy between photons from $Z\gamma$ and
photons from $\gamma\gamma$ final states is less than $1\gev$ when
$m_{\tilde W}\gsim 2\tev$.
An extraordinary energy resolution of $\Delta E/E\lsim 0.1\%$ would
be required to resolve the two lines; otherwise, the photons from $\gamma\gamma$ and
$Z\gamma$ would add together in the same energy bin.

The experimental situation for a monochromatic $\gamma\gamma$
signal looks especially good for discovering sub-TeV 
Winos, which might
be generated from some non-thermal sources~\cite{Wells:2003tf}.  However, the
data pressures discussed above are pointing toward heavier
scalar masses in the hundreds of TeV range. In charged supersymmetry
breaking scenario this correlates with a trans-TeV Wino mass.
An obvious prejudice to have, given these pressures, is that
Winos have mass of about 2.3 TeV such that their relic abundance
is created by a normal thermal freeze-out process.  We know from
the generic relationship of gaugino masses to gravitino/scalar mass that
\beq
m_{\tilde W}\simeq 2.3\tev \Longrightarrow m_{3/2}\simeq \tilde m \simeq 0.8\, {\rm PeV}
\eeq
which is our first indication that the scale of supersymmetry breaking (i.e., gravitino
mass) could be the PeV scale.

{\bf PeV-Scale Supersymmetry from Neutrinos:}
In the Standard Model the right-handed neutrino $\nu^c$ is a pure singlet
under all gauge symmetries. However, in the spirit of
``many sectors,"  which for example the string/M-theory landscape seems to imply, we
suppose that it is unlikely that $\nu^c$ is a pure singlet under all gauge symmetries
of nature. To be specific with an illustration, I will assume
that the $\nu^c$ is charged under a new gauge symmetry
$U(1)'$ in such a way that $LH_u\nu^c$ is not allowed in the superpotential.

The next higher order coupling the $\nu^c$ could have with SM states
is through the non-renormalizable interaction
\beq
W=\frac{\lambda}{M_{Pl}}\phi LH_u\nu^c
\label{W neutrino}
\eeq
where $\phi$ is an exotic field which breaks the $U(1)'$ symmetry when
it condenses, and has the right charge assignment such that
the above operator is allowed.  
When $\phi$ condenses, the Dirac neutrino mass that results is
\beq
m_\nu=\frac{\lambda}{M_{Pl}}\langle \phi H_u\rangle .
\label{mv theory}
\eeq

We know from atmospheric neutrino oscillation experiments that
the participating neutrino masses  must satisfy~\cite{Maltoni:2003da}
\beq
\Delta m_\nu^2 \simeq 10^{-3}\, {\rm eV}^2.
\eeq
This implies that a natural value for the neutrino masses
would be $m_\nu \simeq 0.1\, {\rm eV}$.
If we plug this estimate into eq.~\ref{mv theory} we can
compute the required value of $\langle \phi\rangle$:
\beq
\langle \phi\rangle \simeq \frac{(1\pev)}{\lambda\sin\beta}.
\eeq
Thus, we have another indication of the PeV scale, this time 
from an alternative explanation for neutrino masses.

It does not take much to motivate how $\langle \phi \rangle$ could
be correlated directly with the gravitino and scalar superpartner
masses.  In a many-sectored theory of physics, we assume that 
the symmetry breaking potential for all the sectors is characterized
by the $m_{3/2}$ scale and that our weak scale was one of the
few sectors (or only sector) that happened to have the potential
terms conspire to give a small scale.  If we take this argument
seriously, then at the $m_{3/2}\simeq 1\, {\rm PeV}$ scale
there should be many sectors and many states, only one of
which needs to cooperate to give the neutrino masses.  The PeV scale
would then be very rich physics territory, and could in principle
turn into a target for distant future collider physics programs
if any kind of evidence surfaces for the PeV scale.

{\bf Gauge Coupling Unification:}
Having established that the PeV scale is an interesting one for scalars,
we should check that gauge coupling unification is ok in this scenario.
It is well known that the $\mu$ term value is
crucially important to gauge coupling unification details, and so we
also have some anticipation of possibly restricting $\mu$ from this argument.

Gauge coupling unification is tested by running the dimensionless couplings up to the high scale using
two-loop renormalization group equations and decoupling the contributions of
superpartner states at scales below their mass thresholds.  One needs to know all
the masses of the superpartner states to do this properly.  A dramatic simplification
is to assume all superpartner masses are at a single scale $M_{\rm SUSY}$,
and all superpartners decouple below that scale.  It has been known for some
time now~\cite{Amaldi:1991cn} that if $M_{\rm SUSY}\lsim {\rm few}\, \tev$ the 
gauge couplings unify to within
a percent, well within the range expected of high-scale threshold corrections.

The superpartner masses of PeV-scale supersymmetry are far separated, 
and it is not a good approximation that all superpartner masses should decouple at one scale.
However, there is an effective scale, $M^{\rm eff}_{\rm SUSY}$, which takes into
account the various mass splittings of a 
model~\cite{Langacker:1992rq,Carena:1993ag}.  This scale is introduced for the 
purpose of finding a single scale at which one can decouple all superpartners and
yet retain the complete effect of all the thresholds corrections on the gauge coupling
unification condition.   Just as was the case for $M_{\rm SUSY}$, if $M^{\rm eff}_{\rm SUSY}$
is less than a few TeV gauge coupling unification is fine.

If we assume an anomaly mediated spectrum for the gaugino masses, and
assume that sleptons and squarks and the heavy Higgs boson doublet all
have mass $\sim m_{3/2}$, the value of $M^{\rm eff}_{\rm SUSY}$ is
\beq
M^{\rm eff}_{\rm SUSY}\simeq |\mu| \left( \frac{\alpha_2}{3\alpha_3}\right)^{28/19}\,
\left( \frac{\alpha_2}{4\pi}\right)^{4/19}\, \left( \frac{m_{3/2}}{|\mu|}\right)^{7/19}.
\eeq
Evaluating the  numerical factors and writing it in a more suggestive form, we get
\beq
M^{\rm eff}_{\rm SUSY}\simeq \frac{m_{3/2}}{100}\,
\left( \frac{|\mu|}{m_{3/2}}\right)^{12/19}< 10\tev,
\eeq
where the $10\tev$ number comes from setting $|\mu|$ to its likely maximum value of
$|\mu|\sim m_{3/2}$ and then setting $m_{3/2}$ to its maximum
value of about $1\pev$. 
The numerical value of $M^{\rm eff}_{\rm SUSY}$ is coming out to be
lower than one perhaps would naively expect.  One technical reason
for this is the relatively large ratio of gluino mass to Wino mass.

One might be tempted to be slightly
uncomfortable at the very largest values of $|\mu|\sim m_{3/2}$, where 
$M^{\rm eff}_{\rm SUSY}$ may approach $10\tev$. If insisted upon,
one could require $|\mu|/m_{3/2}\lsim 1/30$ to reduce $M^{\rm eff}_{\rm SUSY}$
below the TeV scale.    
However, $M^{\rm eff}_{\rm SUSY}\simeq 10\tev$  is plenty compatible with 
reasonable grand
unification threshold corrections.
Even best fits to a common superpartner
threshold mass that do not take into account high-scale threshold corrections
allow values as large as $10^4\gev$~\cite{deBoer:2003xm}.  

Therefore, the conclusion of this section is that gauge coupling unification is fine
for any value of $|\mu|$ between the Wino mass and the scalar mass.  That is,
a TeV-scale $M^{\rm eff}_{\rm SUSY}$, which is good for gauge coupling
unification, is  consistent
with the descriptions of the PeV-scale supersymmetry approach discussed above.

{\bf Conclusions:}
PeV-scale supersymmetry solves most of the vexing problems weak-scale supersymmetry
faced (FCNC, CP violation, Higgs mass bound, etc.).  However, it retains all the
good features of low-scale supersymmetry --- dark matter 
and gauge coupling unification in particular.  The only good feature not retained
by PeV-scale supersymmetry is the lack of a Principled Finetuning explanation of the
weak-to-Planck scale hierarchy.  This might be too high of a price to pay for PeV-scale
supersymmetry.  Nevertheless, string/M-theory landscape considerations give us
a plausible reason to dismiss our ordinary intuition of finetuning and follow the
data.  It is noteworthy that generic charged supersymmetry breaking, which gives
rise to loop-suppressed anomaly-mediated gaugino masses, satisfies all the
data pressures on the supersymmetry breaking scale. It is remarkable
that the same PeV numerical
value for the supersymmetry scale can be argued independently from dark matter and neutrino considerations.

Some experimental implications follow from PeV-scale supersymmetry.
First, the Higgs mass is predicted to lie within $125\gev \lsim m_h \lsim 155\gev$, which
can be gleaned by restricting to the $\tilde m\simeq 1\pev$ neighborhood in~\cite{Giudice:2004tc}.
Furthermore, we expect the neutrinos to have Dirac masses generated from non-renormalizable interactions
in the superpotential; the dark matter to be a Wino LSP with mass of about $2.3\pm 0.2\tev$;
and, no deviations from the SM seen by FCNC or CP-violating experiments due to the
superpartners of minimal supersymmetry.  A positive beyond-the-SM experimental signature of this
scenario, which there are precious few in the energy domains of current experiments, would be 
$\tilde W\tilde W\to \gamma\gamma,Z\gamma$ and $e^+X$ annihilations in the galactic halo.

\bigskip
{\bf Acknowledgements:}
This work was supported by the Department of Energy and the Alfred P. Sloan Foundation.
I wish to thank G. Kane, J. Kumar, S. Martin, K. Tobe and M. Toharia for discussions.

\end{document}